# TOPOLOGY PROPERTIES OF WRITTEN HUMAN LANGUAGE

## B. R. Gadjiev and T. B. Progulova

International University for Nature, Society and Man,
19 Universitetskaya Street, 141980 Dubna, Russia, gadjiev@uni-dubna.ru

**Abstract**

We use the extended Barabasi model without the rewired process and show that the degree distribution for the corresponding networks is the Tsallis distribution.

We offer an analysis of the novel "The Sound and the Fury" by W. Faulkner in English and in Russian, and show that the degree distributions of the relevant word networks are described with the Tsallis distribution.

We have constructed degree distributions for each of the relevant word networks and defined the value of the nonextensivity parameter $q$ with the maximum likelihood method. For the novel text in English $q_B = 1.57$; $q_K = 1.49$; $q_J = 1.53$; $q_A = 1.47$; $q_T = 1.54$, and for the translation into Russian $q_B = 1.50$; $q_K = 1.42$; $q_D = 1.46$; $q_A = 1.40$; $q_T = 1.47$. Therefore, if the translation of the novel is regarded as mapping, the nonextensivity parameters ordering $q_B > q_T > q_D > q_K > q_A$ is an invariant of this mapping.

**Introduction**

Written language is one of the most important examples of self-organization systems. It is interesting not only from the linguistic and philosophic points of view, but in the sense that it allows the accomplishment of the full empiric structure research. If we examine a separate piece as a system and represent it as a finite network where words are vertices, with two vertices being connected if they are neighbours, we can analyze the local structure and topology of this network. Moreover, it is possible to simulate the process of the network growth and reveal the self-organization principles of the word network [1].

In many real growing networks the mean number of connections per vertex increases with time. . Such a growth can be called accelerated [2, 3]. It results in the situation when the degree distribution of the word network in the log-log scale has two linear parts with different slopes divided by the crossover point. However, empiric research shows that word networks are grow with acceleration not always.

Fundamental distributions of statistical mechanics for complex systems are derived from general principles; therefore, based on their grounds, the topology description of the word network is of considerable interest. In this paper, we present a model of a growing network that leads to the Tsallis distribution for the degree distribution. Besides, we offer an analysis of the novel "The Sound and the Fury" by W. Faulkner in English and in Russian, and show that the degree distributions of the relevant word networks are described with the Tsallis distribution [4].

**Review of previous papers**

Word networks have been studied in a number of papers [2, 3]. The word network vertices are various words of the language. However, the definition of the words interaction in the network is ambiguous. For example, words-neighbours in the sentence can be regarded interacting, as well as the first and the second neighbours or all words belonging to the same sentence. Nevertheless, studies have shown that various methods of construction bring about word networks with similar topology [3].

In paper [2] the analysis results of the word network which corresponds to the British National Corpus is presented. The British National Corpus is a collection of text samples of the contemporary spoken and written English language. The collection of the texts consists of about 70 million words. The corresponding network has about 470 000 vertices. The authors analyzed in paper [2] the network topology and showed that the degrees distribution for the constructed network has two linear areas in the log-log scale.

The stochastic theory of the human language evolution based on assumption of the language as an evolving network of interacting words is presented in [3]. The following rules for the word network growth are given in this paper:

1. In each discrete moment of time a new vertex – a word – is added to the network. Being added, the new vertex connects to the vertex which is already in the network, according to the principle of preferable attachment.
2. Simultaneously, *ct* of new edges appears among the words existing in the network. Here *c* is a continuous co-efficient that characterizes the network specificity.

In the continuous approximation the equation of motion for the degree *k(s,t)* of the node *s* in the instant *t* is written in the form

$$\frac{\partial k(s,t)}{\partial t} = (1 + 2ct) \frac{k(s,t)}{\int_0^t du\, k(u,t)} \qquad (1)$$

with the initial condition $k(t,t) = 1$.

The solution of this equation has the form:

$$k(s,t) = \left(\frac{t}{s}\right)^{1/2} \left(\frac{2+ct}{2+cs}\right)^{3/2}. \qquad (2)$$

By definition

$$P(k,t) = -\frac{1}{t}\left[\frac{\partial k(s,t)}{\partial s}\right]^{-1}_{s=s(k,t)} \qquad (3)$$

Thus,

$$P(k,t) = -\frac{1}{t}\left[\frac{\partial k(s,t)}{\partial s}\right]^{-1}_{s=s(k,t)} = \frac{1}{t}\frac{s(2+cs)}{1+2cs}\frac{1}{k(s,t)} \qquad (4)$$

As $k(s,t) \sim s^{-\beta}$, it follows from equation (2) that at small $s$ the value $\beta = 1/2$, while at large $s$ $\beta = 2$. Using the scaling ratio $\beta(\gamma - 1) = 1$ we obtain $\gamma = 3$ at large degrees, while at small degrees it is $\gamma = 3/2$.

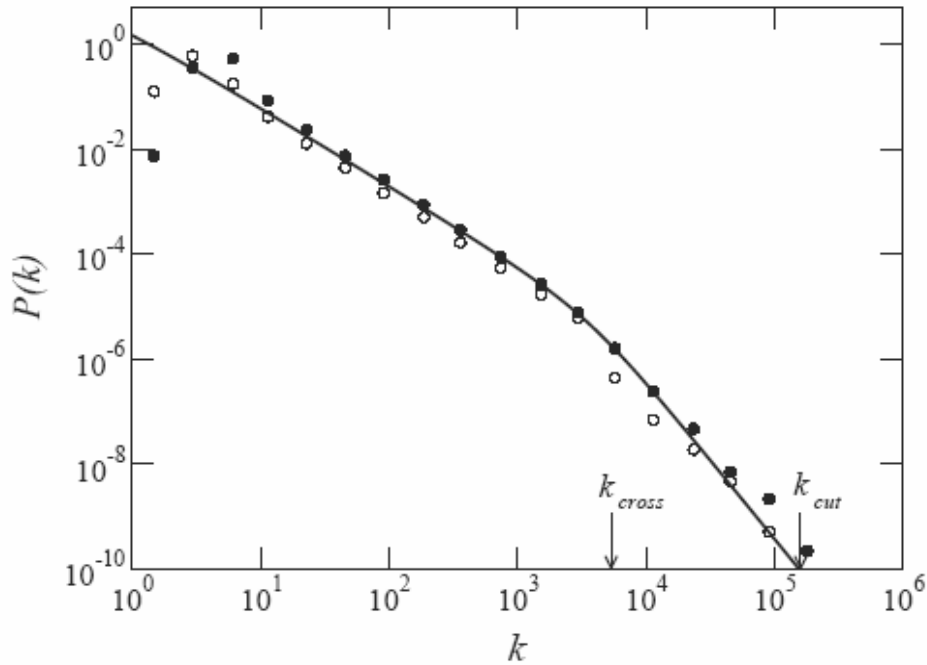

Fig. 1. The empirical distribution of the number of connection of words in the Word Web [3]

Dependence (4) describes sufficiently well the degree distribution of the word network that corresponds to the British National Corpus. However, we intend to demonstrate certain examples of word networks whose degree distribution is not described with equation (4) [5].

**Data analysis.** We present in the paper an analysis of the word network of the novel "The Sound and the Fury" by W. Faulkner, in English and in Russian.

Our special interest in the novel "The Sound and the Fury" by W. Faulkner was the fact that its four parts correspond to four types of perception; however, they are induced by the same events. Therefore, we have analyzed the corresponding parts of the novel and substantively constructed word networks that correspond to various types of perception.

We have constructed the word networks in the following way: the network vertices are various words of the language, and non oriented edges are the connections between interacting words. Finite chains in the graph terms correspond to separate sentences in our representation of the word network. At the network evolution, these finite chains intersect in the vertices that correspond to general words. As a result, a network is formed that is a multigraph. The network in this representation is considered as growing, and the network topology is defined by the degree distribution.

The first network (B) corresponds to the infant, pre-logic, sensuous perception of Benjamin, the second network (K) corresponds to the adolescent broken perception of Quentin, the third one (J) – to the adult, pragmatic, unimaginative perception of Jason and the fourth network (A) corresponds to the wider and more independent perception of the author-observer. Besides, we have analyzed the word network (T) that corresponds to the novel on the whole.

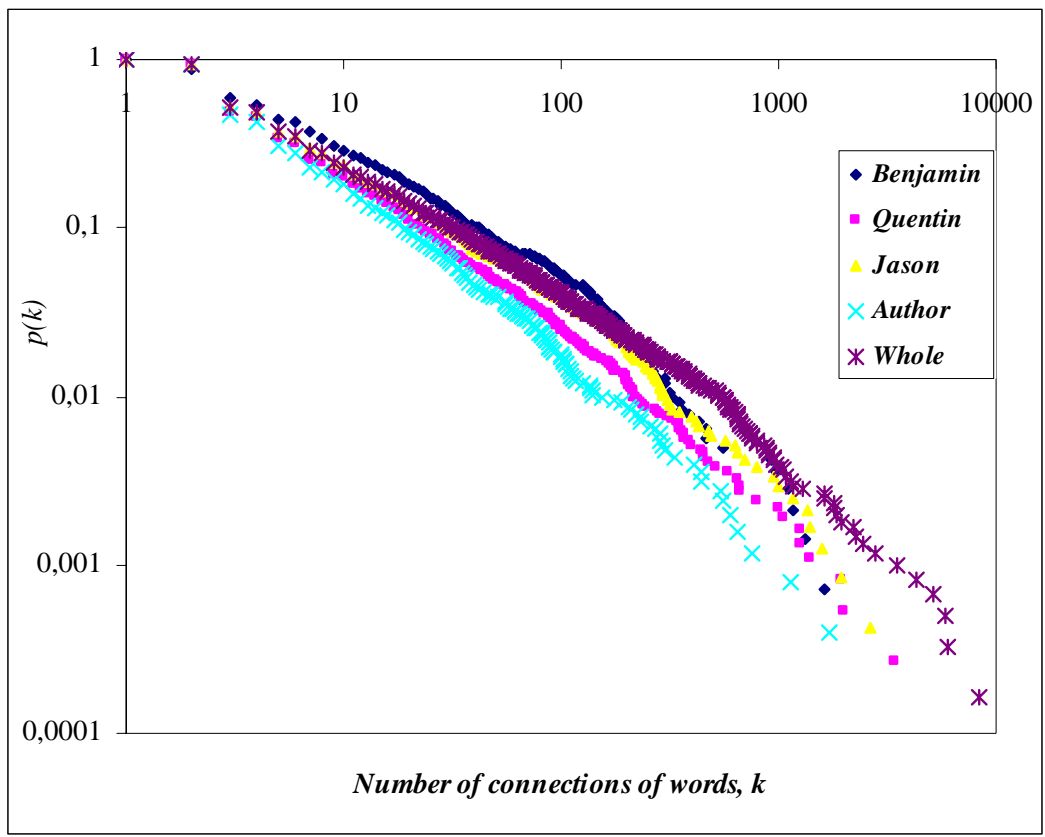

Fig.2. Degree distributions for word networks based on the parts of the novel "The Sound and the Fury" by W.Faulkner on behalf of Benjamin, Quentin, Jason, the author, as well as for the whole text of the novel (in English)

Vertices degree distributions have been constructed for each network. They are given in Fig. 2-3.

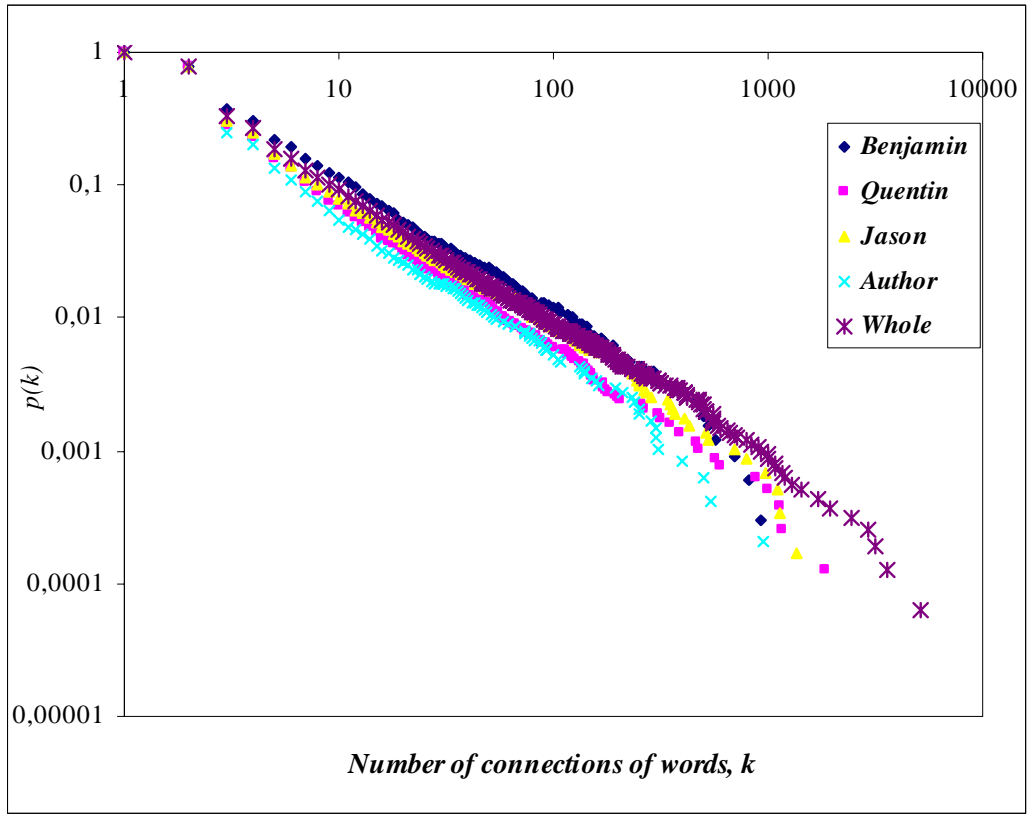

Fig.3. Degree distributions for word networks constructed on the parts of the novel "The Sound and the Fury" by W. Faulkner on behalf of Benjamin, Quentin, Jason, the author, as well as for the whole text of the novel (translated into Russian)

The straight-forward analysis shows that the obtained degree distributions of the word networks are not described with equation (4).

Further, we discuss the variant of the generalized model of the Barabasi growing network and analytically show that such a network has the topology described with the Tsallis distribution.

**Architecture and self-organization of a growing word network.** Let us assume that the growth process starts from the $m_0$ isolated vertices, and one of the following operations is implemented at each time step [1]:

(i) with the $p$ probability we add $m$ $(m \leq m_0)$ of new connections; for each of the latter, we select one vertex in random manner and another one with probability

$$\Pi(k_i) = \frac{k_i + 1}{\sum_j (k_j + 1)}, \quad (5)$$

i.e. a new connection is "preferably" bound to the higher degree vertex.

(ii) with the $1-p$ probability we add the new vertex with new $m$ connections. The connection with the $i$ node that is already in the system is exercised with the $\Pi(k_i)$ probability.

The motion equation for the $k_i$ degree of the $i$ node in the mean-field theory approximation is represented by the expression

$$\frac{\partial k_i}{\partial t} = pm\frac{1}{N} + m\frac{k_i + 1}{\sum_j (k_j + 1)} \quad (6)$$

In (6) the dimension of the $N$ system and the total number of connections $\sum_j k_j$ change with time as $N(t) = m_0 + (1-p)t$, and $\sum_j k_j = 2mt - m$. Hence, at large $t$, neglecting the $m_0$ and $m$ constants, we obtain $N(t) \cong (1-p)t$ and $\sum_j k_j = 2mt$. Consequently, equation (6) becomes

$$\frac{\partial k_i}{\partial t} = pm\frac{1}{(1-p)t} + \frac{m(k_i + 1)}{2t + (1-p)t} \quad (7)$$

Having introduced the symbols $\phi(p,m) = p(2m/(1-p) + 1)$ and $\theta(p,m) = (2m + 1 - p)/m$, we write equation (7) in the following way:

$$\frac{\partial k_i}{\partial t} = (\phi(p,m) + 1 + k_i)\frac{1}{\theta(p,m)t} \quad (8)$$

Accounting for the initial hypothesis $k_i(t_i) = m$ we find the solution for equation (8)

$$k_i(t) = (\phi(p,m) + m + 1)\left(\frac{t}{t_i}\right)^{\frac{1}{\theta(p,m)}} - \phi(p,m) - 1 \quad (9)$$

Let us define the $P[k_i(t) < k]$ probability of the notion that the $i$ node has the degree $k_i(t)$, which is less than $k$. Using (9),

$$P[k_i(t) < k] = P[t_i > \psi(p,q,m)t],$$

where

$$\psi(p,m) = \frac{(\phi(p,m) + m + 1)^{\theta(p,m)}}{(\phi(p,m) + k + 1)^{\theta(p,m)}}.$$

As $0 \leq t_i \leq t$, $0 < \psi(p,m) < 1$.

As $P_i(t_i) = \dfrac{1}{m_0 + t}$, for the cumulative distribution we have

$$P[k_i(t) < k] = 1 - \frac{\psi(p,m)t}{m_0 + t}. \tag{10}$$

Then, defining the degree distribution $P(k) = \dfrac{\partial P[k_i(t) < k]}{\partial k}$, we get

$$P(k,t) = \frac{t}{m_0 + t} \theta(p,m)(\phi(p,m) + m + 1)^{\theta(p,m)} (\phi(p,m) + k + 1)^{-\theta(p,m)-1}, \tag{11}$$

and, introducing the symbols $\tau(p,m) = \theta(p,m)[m + \phi(p,m+1)]^{\theta(p,m)}$, $\kappa(p,m) = \phi(p,m) + 1$, $\gamma(p,m) = \theta(p,m) + 1$, we are able to write

$$P(k,t) = \frac{t}{m_0 + t} \tau(p,m)(\kappa(p,m) + k)^{-\gamma(p,m)} \tag{12}$$

The statistical function of the degree distribution $P(k) = \lim_{t \to \infty} P(k,t)$ is expressed in

$$P(k,t) = \tau(p,m)(\kappa(p,m) + k)^{-\gamma(p,m)}. \tag{13}$$

Let us introduce the symbols $\dfrac{1}{1-q} = -\gamma$, $\dfrac{1}{\kappa(p,m)} = -(1-q)\beta$. Then $\beta = \dfrac{\gamma(p,m)}{\kappa(p,m)}$, and from equation (13) it follows that $P(k) = \tau(p,m)\kappa(p,m)^{-\gamma}(1 - (1-q)\beta k)^{\frac{1}{1-q}}$

Using the normalization hypothesis $P(k)$ we obtain

$$P(k) = \frac{1}{Z}(1 - (1-q)\beta k)^{\frac{1}{1-q}}, \tag{14}$$

where $Z = \sum_k (1 - (1-q)\beta k)^{\frac{1}{1-q}}$. The distribution function is the nonextensive Tsallis distribution.

The value of the nonextensivity parameter $q$ has been determined for each network with the maximum likelihood method. For the novel text in English $q_B = 1.57$; $q_K = 1.49$; $q_J = 1.53$; $q_A = 1.47$; $q_T = 1.54$, and for the translation into Russian $q_B = 1.50$; $q_K = 1.42$; $q_D = 1.46$; $q_A = 1.40$; $q_T = 1.47$. Therefore if the translation of the novel is regarded as mapping, the nonextensivity parameters ordering $q_B > q_T > q_D > q_K > q_A$ is an invariant of this mapping. We present examples of word networks for which the accelerated growth is not characteristic.

**Conclusion.** Human language is very important for general research in network theory as, firstly, the data are accessible and precise, and secondly, due to the detailed knowledge of its local organizational rules. We have analyzed in the present paper the topology of the written language through the representation in word network. We postulate a model that allows us to describe the empiric behaviour of the word network. We show that the word network degree distribution is described by the Tsallis statistics. We emphasize that the mechanisms which lead to the Tsallis statistics can be different. Besides, the Tsallis distribution can be successfully applied to the description of the topology of networks that not show accelerated growth and show accelerated growth.

The given analysis of the word network based on the novel "The Sound and the Fury" by W. Faulkner in English and in Russian is of interest in the aspect that its four parts correspond to four types of perception but are induced by the same events. We have constructed degree distributions for each of the relevant word networks and defined the value of the nonextensivity parameter $q$ with the maximum likelihood method. For the novel text in English $q_B = 1.57$; $q_K = 1.49$; $q_J = 1.53$; $q_A = 1.47$; $q_T = 1.54$, and for the translation into Russian $q_B = 1.50$; $q_K = 1.42$; $q_D = 1.46$; $q_A = 1.40$; $q_T = 1.47$. If the translation of the novel is regarded as mapping, the nonextensivity parameters ordering $q_B > q_T > q_D > q_K > q_A$ is an invariant of this mapping.

**References**

1. Albert R. and Barabasi A.-L. Topology of Evolving Networks: Local Events and Universality. Phys. Rev. Lett. 85 (24), 2000, 5234-5237
2. Ferrer Cancho R. and Sole R.V. The small-world of human language. Proc. R. Soc. Lond. B. **268**, 2261.
3. Dorogovtsev S.N. and Mendes J.F.F. Language as an evolving Word Web. Proc. Royal Soc. London B. 2001. **268.** 2603-2606.
4. Tsallis C. Nonextensive statistics: theoretical, experimental and computational evidences and connections. Braz. J. Phys. **29**, 1999. P. 1-35.
5. Gadjiev B. R., Progulova T. B., Kalinkina E. A. Self-organization of word network, Proceedings of the International Conference "Advanced Synergetics", Tver: Tver State University, p. 322-326.